# Linear, Machine Learning and Probabilistic Approaches for Time Series Analysis

B.M.Pavlyshenko
Ivan Franko National University of Lviv, e-mail: b.pavlyshenko@gmail.com

*In this paper we study different approaches for time series modeling. The forecasting approaches using linear models, ARIMA alpgorithm, XGBoost machine learning algorithm are described. Results of different model combinations are shown. For probabilistic modeling the approaches using copulas and Bayesian inference are considered.*

## I. INTRODUCTION

Time series analysis, especially forecasting, is an important problem of modern predictive analytics. The goal of this study is to consider different aproaches for time series modeling. For our analysis, we used stores sales historical data from Kaggle competition "Rossmann Store Sales" [1]. These data represent the sales time series of Rossmann stores. For time series forecasting such approaches as linear models and ARIMA algorithm are widely used. Machine lerning algorithm make it possible to find patterns in the time series. Sometimes we need to forecast not only more probable values of sales but also their distribution. Especially we need it in the risk analysis for assessing different risks related to sales dynamics. In this case, we need to take into account sales distributions and dependencies between sales time series features (e.g. day of week, month, average sales, etc.) and external factors such as promo, distance to competitors, etc. One can consider sales as a stochastic variable with some marginal distributions. If we have sales distribution, we can calculate value at risk (VaR) which is one of risk assessment features. In probabilistic analysis of sales, we can use copulas which allows us to analyze the dependence between sales and different factors. To find distributions of model parameters Bayesian inference approach can be used.

## II. LINEAR MODELS AND MACHINE LEARNING APPROACHES

For our analysis, we used stores sales historical data. To compare different forecasting approaches we used two last months of the historical data as validation data for accuracy scoring using root mean squared error (RMSE). For the comparison, we used the following methods: ARIMA using R package "forecast" [2], linear regression with LASSO regularization using R package "lars" [3], conditional inference trees with linear regression on the leaf using mob() function from R package "party" [4], gradient boosting XGBoost model using R package "xgboost" [5]. Package "xgboost" (short term for eXtreme Gradient Boosting) is an efficient and scalable implementation of gradient boosting framework [6,7]. The package includes efficient linear model solver and tree learning algorithm. We also used combined approaches such as linear blending ARIMA and gradient boosting model, stacking with the use of linear regression on the first step and gradient boosting on the second step. We used two ways of classifications – the first way is based on the time series approach and the second one is based on the independent and identically distributed variables. We consider sales in the natural logarithmic scale. Figure 1 shows typical time series of store sales.

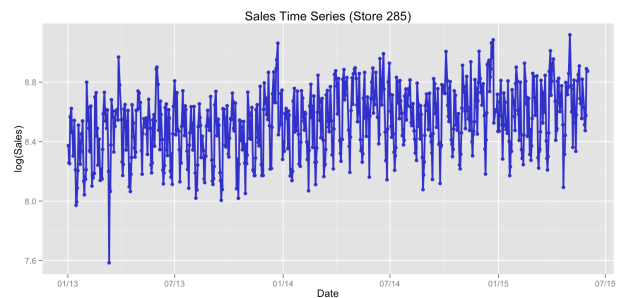
Figure 1. Typical time series of store sales.

Figure 2 shows the example of time series forecasting by different methods with RMSE error metric. Let us consider the case of time series forecasting using linear blending of ARIMA and XGBoost models. For arbitrary chosen store (Store 285) we received RMSE=0.11 for ARIMA model, RMSE=0.107 for XGBoost model and RMSE=0.093 for linear blending of ARIMA and XGBoost models. Let us consider the case of using stacking with linear regression on the first step and xgboost on the second step. For arbitrary chosen store (Store 95) we received RMSE=0.122 for XGBoost model and RMSE=0.117 for stacking model.

We also studied the case of time series forecasting using XGBoost model with time series approach and xgboost model based on independent and identically distributed variables. For arbitrary chosen store (Store 95) we received RMSE=0.138 for XGBoost model with time series approach and RMSE=0.118 for XGBoost model with i.i.d approach. The obtained results show that for different stores the best accuracy is released by different approaches. For each type of time series, we may develop an optimized approach which can be based on the combination of different predictive models.

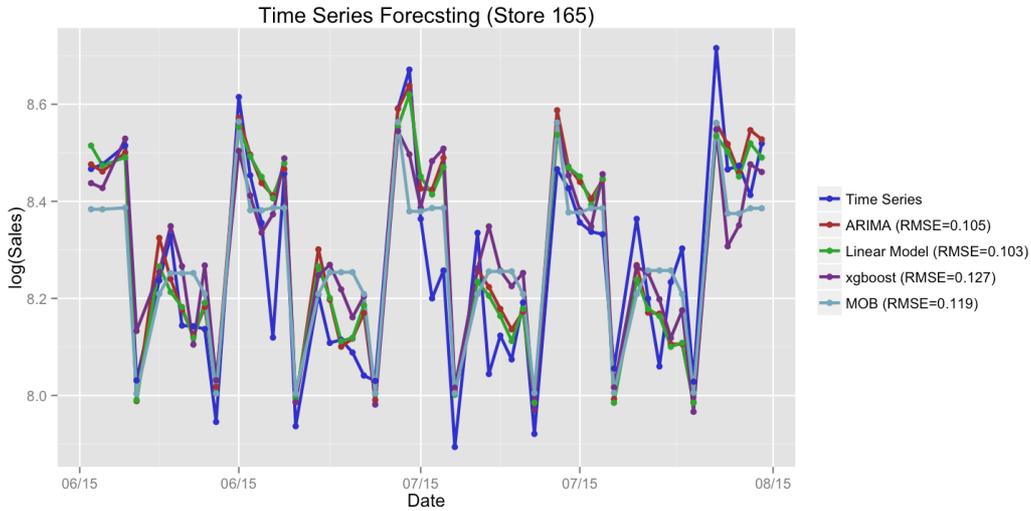

Figure 2. Time series forecastings by different methods.

III. COPULAS APPROACH FOR MODELING

A copula is a multivariate probability distribution for which the marginal probability distribution of each variable is uniform. Copulas are used to describe the dependence between random variables. Sklar's Theorem states that any multivariate joint distribution can be written in terms of univariate marginal distribution functions and a copula, which describes the dependence structure between the variables. The copula contains all information on the dependence structure between the variables, whereas the marginal cumulative distribution functions contain all information on the marginal distributions. For the case study, we use the same sales time series, which represent sales in the stores network. We used "copula" R package [8] for modeling. We consider sales in the natural logarithmic scale. For our analysis, we take such features as sales (variable logSales), previous day sales (variable prevLogSales), number of customers that visited a store (variable Customers). First of all we take one sales time series for one arbitrary store. Marginal distributions and dependencies with correlation coefficient are shown on the figure 3. On the figures 4,5 the pseudo observations of investigated features are shown. These figures represent stochastic dependencies of investigated variable. Our next objective is to find such copulas, which will approximate these dependencies. We chose t-copula for modeling. Using maximum likelihood method, one can find fitting parameters for copula. The probability density function for calculated fitted t-copula is shown on the figure 6.

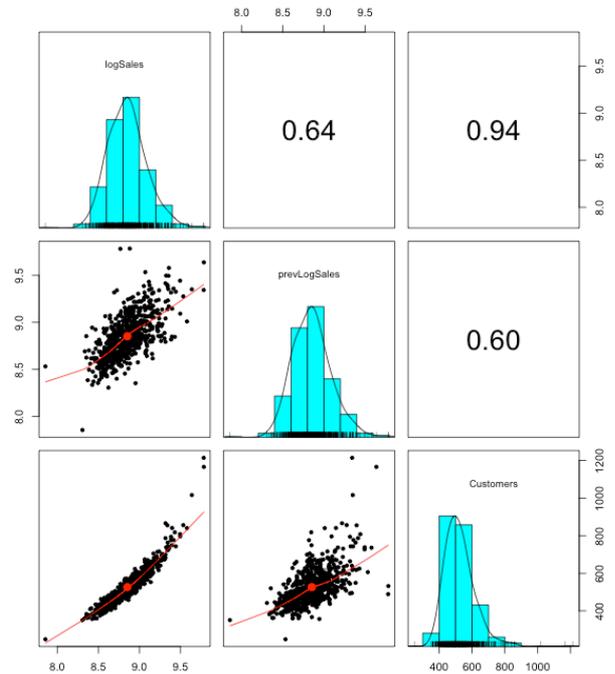

Figure 3. Marginal distributions and correlation coefficient.

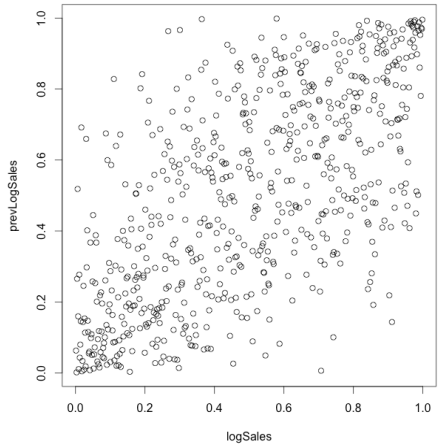

Figure 4. Pseudo observations for logSales and prevLogSales variables.

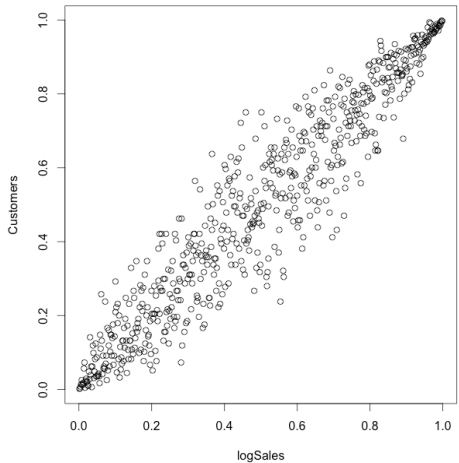

Figure 5. Pseudo observations for logSales and Customers variables.

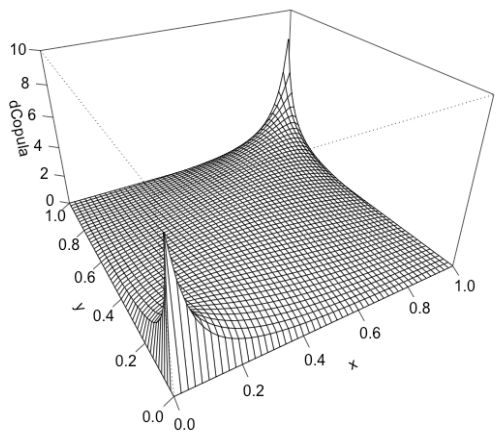

Figure 6. The probability density function for fitted t-copula.

Having fitted copula and marginal distributions, we can generate pseudo-random samples of investigated variables by applying inverse marginal comulative distribution function (CDF) to each dimensional variable of fitted copula. To construct multivariate distribution of dependent variables, we chose gamma distribution for marginal distributions of logSales and Customers variables. Having fitted copula and finding parameters for these gamma distributions from historical data, we generate pseudo-random samples with the probability density function (PDF), shown on the figure 7.

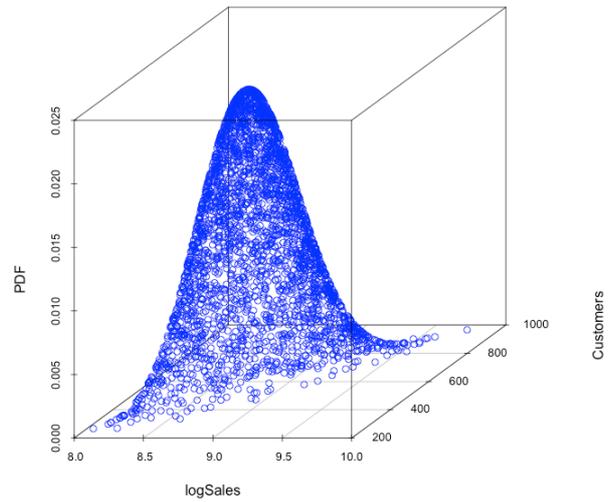

Figure 7. The PDF of generated pseudo-random samples using fitted t-copula and marginal distributions.

If we need to analyze multivariate dependences with more than two variables, it is effective to use vine copulas, which enable us to construct complex multivariate copula using bivariate ones. For studying vine copula, we used CDVine R package [9]. Let us consider such features of sales time series as sales (variable logSales), mean sales per day for store (variable meanLogSales) and promo action (variable Promo). In this case, we analyze sales for stores chain. To analyze the stochastic dependence we used canonical vine copula. First tree for fitted canonical vine copula with Kendall's tau values is shown on the figure 8.

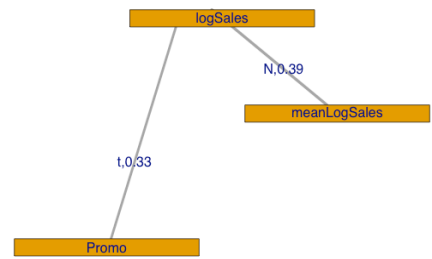

Figure 8. The tree for fitted canonical vine copula with Kendall's tau values.

We chose a t-copula for the logSales-Promo dependency and a normal for the logSales-meanLogSales dependency. The pseudo observations for constructed canonical vine copula are shown on the figure 9 in the dimension of logSales and meanLogSales .

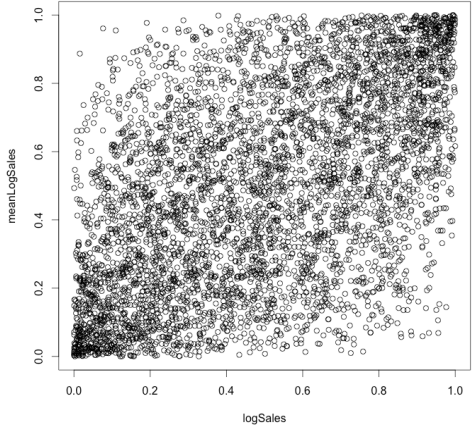

Figure 9. Pseudo-observations for constructed canonical vine copula.

As the case study shows, the use of copula make it possible to model stochastic dependencies between different factors of sales time series separately from their marginal distributions. This can be considered as an additional approach in the sales time series analysis.

## IV. BAYESIAN INFERENCE

For Bayesian inference case study, we take such features as promo, seasonality factors (week day, month day, month of year). As well as in the previous studies, we consider sales in the natural logarithmic scale. As the example we take one sales time series for one arbitrary store. For Bayesian inference, we used Markov Chain Monte Carlo (MCMC) algorithm from MCMCpack R package [10]. For time series modeling, we used the linear regression with Gaussian errors. Trace plots of samples vs the simulation index can be very useful in assessing convergence. The trace plot for promo coefficient is shown on the figure 10.

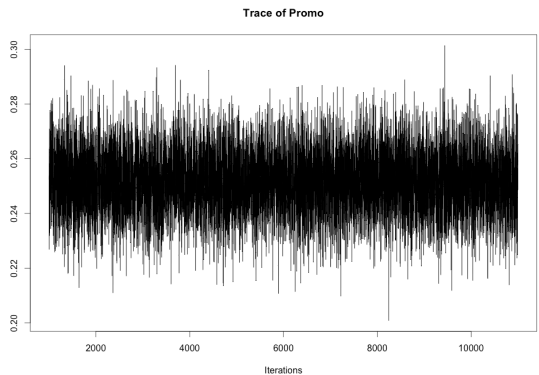

Figure 10. Trace plot for promo coefficient.

The trace plot demonstrates the stationary process, which means good convergence and sufficient burn-in period in the MCMC algorithm. The similar trace plots were received for other coefficients in the linear regression. The density of distributions of some regression coefficients for chosen arbitrary store are shown on the figure 11.

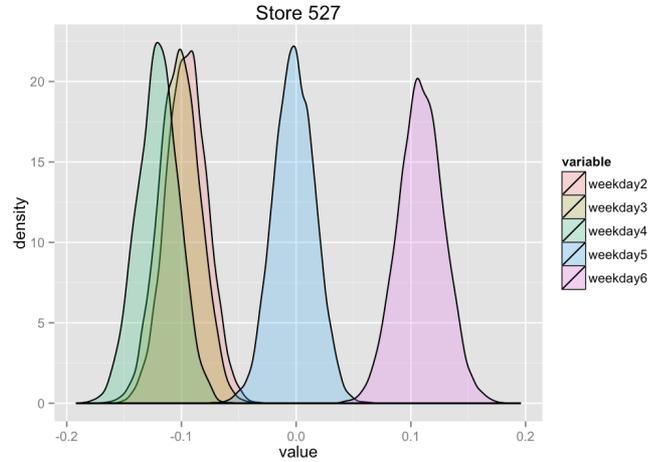

Figure 11. Density of distributions of regression coefficients.

The figure 12 shows the examples for box plots for some regression coefficients

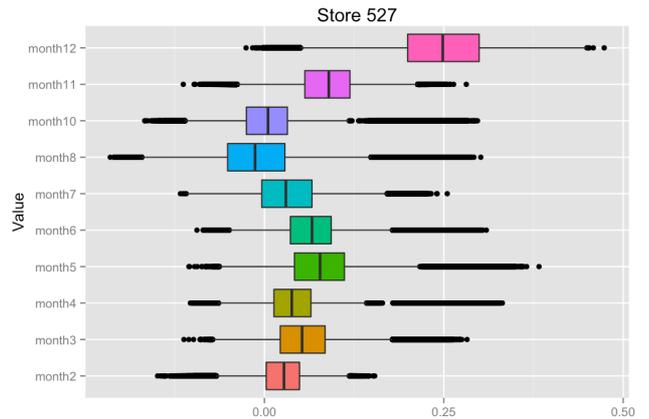

Figure 12. Box plots for regression coefficients.

Sales time series can have outliers and it is important to take into account this fact using heavy tails distributions instead of Gaussian distribution. For the linear regression with variables with different type of distributions we used Bayesian hierarchical model. We conducted the case study using JAGS sampler [11] software with "rjags" R package. For modeling, we take into account mean sales for the store, sales, and promo. We consider sales and mean sales for the store in the natural logarithmic scale. For mean sales for stores, we used Gaussian distribution, for sales – Student distribution, and for promo – Bernoulli distribution. In this model, we consider sales as an independent and identically distributed random variable without separating sales for different stores. Info about the store is represented by mean sales for the store values. Mean sales for the store (variable meanLogSales) vs sales (variable logSales) obtained for considered Bayesian model are shown on the figure 13.

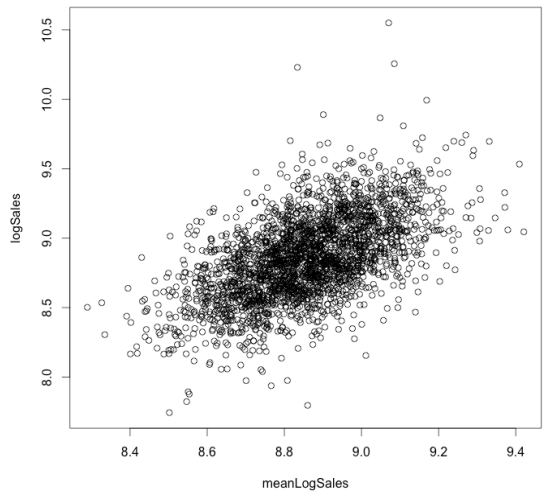

Figure 13. Mean sales for the store (variable meanLogSales) vs sales (variable logSales) obtained for fitted Bayesian model.

As the case study shows, the use of Bayesian approach allows us to model stochastic dependencies between different factors of sales time series and receive the distributions for model parameters. Such an approach can be useful for assessing different risks related to sales dynamics.

## V. CONCLUSION

In our cases study we showed different approaches for time series modeling. Forecasting with using linear models, ARIMA algorithm, xgboost machine learning algorithm are described. The results of different models combinations are shown. For probabilistic modeling, the approach with using copulas is shown. The Bayesian inference was applied for time series linear regression case. For time series forecasting the different models combinations technics can give better RMSE accuracy comparing to single algorithms. The probabilistic approach for time series modeling is important in the risk assessment problems. The copula approach gives one the ability to model probabilistic dependence between target values and extreme factors which is useful when a target variable has non Gausian probability density function with heavy tails. Bayesian models can be used to find distributions of coefficients in the linear model of time series. Having model parameters distributions one can find the distribution of target values using Monte-Carlo approaches. For each type of time series, one can develop an optimized approach, which can be based on the combination of different predictive models.